\documentstyle[epsf,prb,aps]{revtex}
\begin{document}

\title{Autocorrelations from the transfer matrix DMRG method}

\author{F.Naef, X.Wang and X.Zotos}
\address{
Institut Romand de Recherche Num\'erique en Physique des
Mat\'eriaux (IRRMA), \\
INR-Ecublens, CH-1015 Lausanne, Switzerland}

\author{W. von der Linden}
\address{
Institut f\"ur Theoretische Physik der Technische Universit\"at
Graz\\ Petersgasse 16, A-8010 Graz, Austria}

\maketitle

\pacs{75.10.Jm, 02.70.-c, 75.40.Gb, 76.60.-k}

\begin{abstract}
Extending the transfer matrix DMRG algorithm, we are able to calculate
imaginary time spin autocorrelations with high accuracy 
(absolute error $<10^{-6}$) over a wide temperature range 
($0<\beta J<20$). 
After analytic continuation using the rules of
probability theory along with the entropic prior (MaxEnt), we
obtain real frequency spectra for the XY model, the isotropic
Heisenberg and the gaped Heisenberg-Ising model.
Available exact results in some limits allow for a
critical evaluation of the quality of answers expected from this procedure. 
We find that high precision data are still insufficient for resolving
specific lineshapes such as low frequency divergences. However, the
method is appropriate for identifying low temperature gaps and peak
positions.
\end{abstract}

\section{Introduction}
Finite temperature dynamic correlations are the link between experiment and 
theory. In spite of their relevance, it seems that for strongly 
interacting electronic or magnetic systems, no reliable, direct theoretical 
methods exist for their evaluation. 
This absence is particularly felt recently in the field of quasi 
one-dimensional magnetic materials, where excellent samples, detailed 
neutron scattering and NMR experiments\cite{tak,imai} call for a better 
understanding of dynamic form factors.
In this context, numerical simulation techniques can provide valuable information.
However, these are also subject to rather severe limitations. 
On the one hand, real time methods
based on the exact diagonalization of the Hamiltonian matrix
have been restricted to small systems. As a consequence,
the extrapolation to the thermodynamic limit is often unreliable,
especially in the low temperature regime.\cite{petlan} Related
methods such as the moment expansion or the recursion method\cite{vismuell}
can provide reliable short time correlations, however, the extrapolation
to long times is left uncertain. Until now, real time methods
have been insufficient for discussing low frequency properties,
for instance, they cannot decide about diffusive or ballistic transport.
\cite{zn,bl,fab} 

On the other hand, imaginary time methods such as quantum Monte Carlo
\cite{sss} (QMC)
have to face the ill-conditioned analytical continuation problem.
In this context, it is reasonable to expect that the higher accuracy
of the transfer matrix DMRG calculations,\cite{bursill,wang} free of 
stochastic errors, might help. This method has been applied successfully 
to several interesting 
(quasi-)one dimensional systems.
\cite{Shibata,SATSU,shibata,Xiang,Rommer,sch2,sch3,wx,klu}
In this work we describe in detail an extension of the transfer matrix DMRG
method proposed in Ref.\onlinecite{wang} to calculate 
imaginary time correlations.\cite{japs} 
We show that these can be very accurately determined, especially for 
local correlations. 
In addition, we present an extended critical discussion on the potential 
of this method to obtain reliable real frequency spectra. 
Of course, the transfer matrix DMRG method is limited to 
quasi-one dimensional systems.

In order to investigate whether the combination of transfer matrix
DMRG and analytical
continuation methods can provide accurate dynamic correlations, 
we focus on the $s=1/2$ antiferromagnetic Heisenberg Hamiltonian: 
\begin{equation}
H=J \sum_{l} (S_l^x S_{l+1}^x +
S_l^y S_{l+1}^y + \Delta S_l^z S_{l+1}^z) ,
\label{heisham}
\end{equation}
where $S_l^{\alpha}=\frac{1}{2}\sigma_l^{\alpha}$,
$\sigma_l^{\alpha}$ are the Pauli spin operators with 
components $\alpha=x,y,z$ at site $l$. In the following we will take $J$ as 
the unit of energy.

This model represents a good playground for tests since in the XY limit
($\Delta=0$) exact results at all temperatures for the longitudinal 
zz-correlation and at $T=0,\infty$ for the transverse xx-correlation are known. 
For $\Delta=1$, and $\Delta\geq 1$ where the spectrum is gaped, the two spinon 
contribution to the transverse correlation at $T=0$ was recently exactly 
evaluated.\cite{karbach,bougou}

The paper is organized as follows: section \ref{s2} explains how to extend the
transfer matrix DMRG technique for obtaining imaginary 
time correlations and also
provides a summary of the analytical continuation methods. In \ref{sxyzz}
and \ref{sxyxx},
we look closely at the XY model and compare our results to exact solutions.
Among various analytical continuation procedures, we find the MaxEnt 
method to be the most reliable. The results 
for the isotropic point ($\Delta=1$) are presented in section \ref{sheis} and 
those for the gaped regime ($\Delta=2,4$) in section \ref{sgap}. 

\section{Technique}
\label{s2}
\subsection{Notation}

Before explaining the transfer matrix DMRG technique, let us first fix some notation.
The basic quantity from which the nuclear spin relaxation rate $1/T_1$ and
the neutron scattering cross section can be determined is the dynamical
structure factor 

\begin{equation}
S^{\alpha}_{ij}(\omega)=\int_{-\infty}^{+\infty} e^{i\omega t}
\langle S^\alpha_i(t)S^\alpha_j(0)\rangle dt
\end{equation}
where $\alpha=z$ for the longitudinal and
$\alpha=x$ for the transverse correlations.
$S^\alpha_i(t)=e^{itH} S^\alpha_i e^{-itH}$ and the average
$\langle\bullet\rangle$ is taken in the canonical ensemble at 
the inverse temperature $\beta=1/T$. $S^{\alpha}_{ij}(\omega)$ is
a positive function and the autocorrelations $i=j$ satisfy
the sum-rule
\begin{equation}
\frac{1}{2\pi}\int_{-\infty}^{+\infty} S^{\alpha}_{ii}(\omega) d\omega=
\frac{1}{4}
\end{equation}
at all temperatures.
Using the transfer matrix DMRG method we will study the imaginary time Green's
function
\begin{equation}
G^{\alpha}_{ij}(\tau)=\langle S^\alpha_i(\tau)S^\alpha_j\rangle
\label{itgf}
\end{equation}
where $S^\alpha_i(\tau)=e^{\tau H} S^\alpha_i e^{-\tau H}$.

It is related to the real frequency correlations through the linear
integral equation:

\begin{eqnarray}
&&G^{\alpha}_{ij}(\tau)=\frac{1}{2\pi}\int_0^{\infty} K(\tau,\omega) 
S^{\alpha}_{ij}(\omega) d\omega\label{intequ0}\\
&&K(\tau,\omega)=e^{-\tau\omega}+e^{-(\beta-\tau)\omega}
\end{eqnarray}
where the detailed balance condition $S^{\alpha}_{ij}(-\omega)=e^{-\beta \omega}
S^{\alpha}_{ij}(\omega)$ has been included in the kernel
$K(\tau,\omega)$. Notice also that using the 
symmetry $K(\tau,\omega)=K(\beta-\tau,\omega)$ it is only necessary to 
calculate $G^{\alpha}_{ij}(\tau)$ in the interval $[0,\beta/2]$. The 
decomposition of the integrand in Eq. (\ref{intequ0}) 
into $K(\tau,\omega)$ and
$S^{\alpha}_{ij}(\omega)$ is not unique. In fact, it is sometimes better 
(see the xx correlations in the XY model) to reconstruct the symmetric
correlation function ${\tilde S}^\alpha_{ij}(\omega)$, even in $\omega$, 
corresponding to the Fourier transform of the anti-
commutator correlations. For this
purpose, we rewrite Eq. (\ref{intequ0}) as 
\begin{equation}
G^{\alpha}_{ij}(\tau)=\frac{1}{2\pi}\int_0^{\infty}
{\tilde K}(\tau,\omega){\tilde S}^\alpha_{ij}(\omega) d\omega
\label{intEq.ym}
\end{equation}
with
\begin{eqnarray}
&&{\tilde K}(\tau,\omega)=K(\tau,\omega)/(1+e^{-\beta\omega}),\nonumber\\
&&{\tilde S}^\alpha_{ij}(\omega)=S^{\alpha}_{ij}(\omega) (1+e^{-\beta\omega}).
\label{nss}
\end{eqnarray}
This scheme will be referred to as the symmetric scheme.

\subsection{Transfer matrix DMRG}
In this purely technical part, we will explain in detail
how one can calculate imaginary time correlations by
extending the transfer matrix DMRG method developed in 
reference[\CITE{wang}].

\subsubsection{Thermodynamics}

\begin{figure}
\epsfxsize=12cm
\epsffile{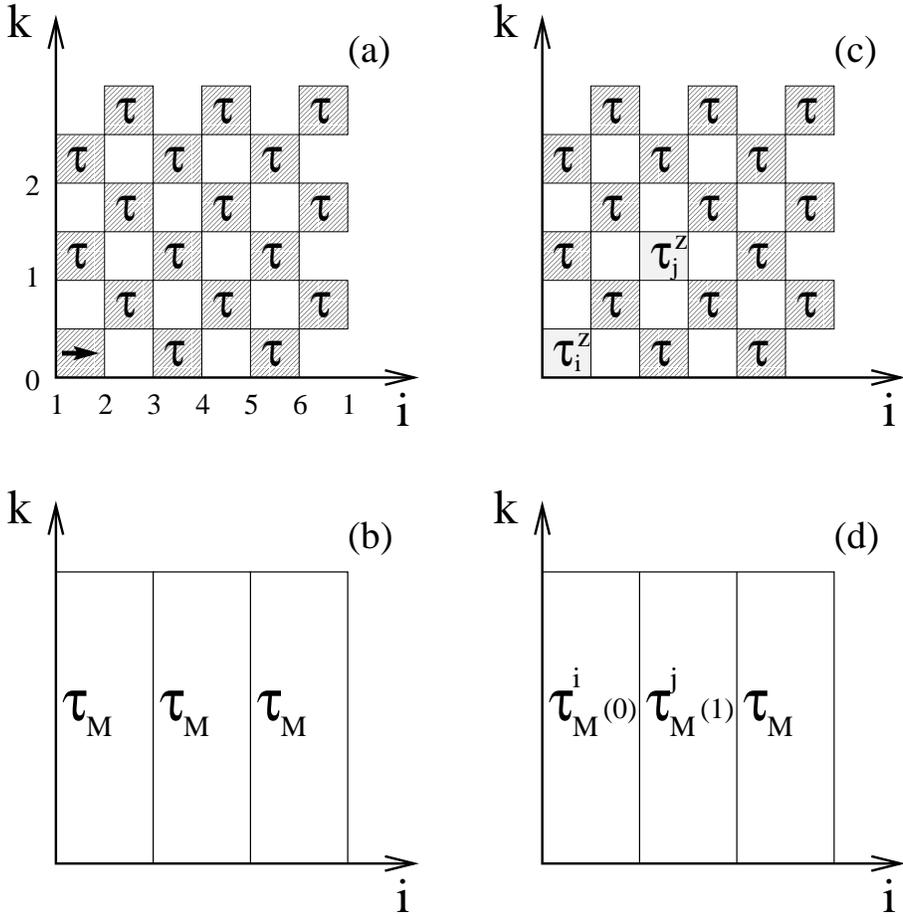}
\caption{The checkerboard representation for $M=3$, $N=6$.
(a) and (b) represent Eqs. (\ref{partfct}) and (\ref{tmdef}),
(c) and (d) Eqs. (\ref{Nij2}) and (\ref{Tmik}) for $i=1$, $j=3,4$.}
\label{checkb}
\end{figure}

Let us briefly recall how the transfer matrix representation of the 
partition function is used to calculate thermodynamic 
quantities. First, we split the Hamiltonian (\ref{heisham}) into
odd and even bond terms:
\begin{eqnarray}
&&H= H_o+ H_e \nonumber\\
&&H_o= h_1 + h_3+ h_5+\cdots \nonumber\\
&&H_e= h_2+ h_4+ h_6+\cdots \nonumber\\
&&h_i= S_i^x S_{i+1}^x+ S_i^y S_{i+1}^y
+\Delta S_i^z S_{i+1}^z , \nonumber
\end{eqnarray}
so that the all even (respectively odd) terms $h_i$ commute with each
other. Then, using the Trotter-Suzuki decomposition 
and doing the standard insertion of the identity 
$1=\prod_{i=1}^N\sum_{\{s_{k'}^i\}}
|s_{k'}^i \rangle\langle s_{k'}^i|$
at each inverse temperature slice $k'=1,2,\ldots,2M$,
we can express the 
partition function in terms of the quantum transfer matrix ${\cal T}_M$:
\cite{Suzuk,Betsu}
\begin{eqnarray}
&&Z={\rm Tr} \, e^{-\beta H}={\rm Tr}\left[ e^{-\epsilon H_o}
e^{-\epsilon H_e}\right]^M+{\cal O}(\epsilon^2)\, ,\nonumber\\
&&{\rm Tr}\left[ e^{-\epsilon H_o} e^{-\epsilon H_e}\right]^M
={\rm Tr}\, \left[{\cal T}_M \right]^{N/2}.
\label{partfct}
\end{eqnarray}
Here $M$ is the Trotter number, $\epsilon=\beta/M$ and $\beta=1/T$.
In the last step of Eq.(\ref{partfct}), the 
summation indices have been
permuted so that the space and Trotter (inverse temperature) directions 
are interchanged. ${\cal T}_M$ is a non-symmetric matrix given by 
the product of $2M$ local transfer matrices
\begin{equation}
\langle \sigma^1_1\cdots\sigma^1_{2M}| 
{\cal T}_M |\sigma_1^3\cdots\sigma^3_{2M} \rangle \label{transf}=
\sum_{\{\sigma^2_k\}} \prod_{k=1}^M \tau (\sigma^1_{2k-1}\sigma^1_{2k} 
|\sigma^2_{2k-1}\sigma^2_{2k})
\tau(\sigma^2_{2k}\sigma^2_{2k+1}|\sigma^3_{2k}\sigma^3_{2k+1})\label{tmdef}
\end{equation}
with periodic boundary condition in the Trotter direction 
($\sigma_{2M+1}=\sigma_1$), $\sigma^i_k=(-1)^{i+k}s^i_k$ and
\begin{equation}
\tau(\sigma^{i}_k \sigma^{i}_{k+1} |\sigma^{i+1}_k\sigma^{i+1}_{k+1})
=\langle s^{i+1}_{k+1}, s^i_{k+1}| 
\exp(-\epsilon \hat h_i) |s^i_k,s^{i+1}_k\rangle. \label{loctm}
\end{equation}
The real space indices are denoted by $i$, the Trotter direction by $k$ and
$S_i^z|s^i_k\rangle=s^i_k |s^i_k\rangle$.
The sign in the definition of $\sigma^{i}_k$ is chosen such that $\tau$ 
conserves $\sigma$: $\sigma^{i}_k+\sigma^{i}_{k+1}=
\sigma^{i+1}_k+\sigma^{i+1}_{k+1}$.
Eqs. (\ref{partfct})-(\ref{loctm}) 
are best represented as a 2-dimensional checkerboard shown in  
Fig. \ref{checkb}(a). The arrow in the
left corner emphasizes that $\tau$ propagates in the
real space direction. 
Fig. \ref{checkb}(b) shows how the full checkerboard is cut along
the real space direction in $N/2$ identical quantum transfer
matrices ${\cal T}_M$.

In the limit $N\rightarrow \infty$,
the partition function $Z$ is given by the maximum 
eigenvalue $\lambda_{\rm max}^{N/2}$.
Thermodynamic quantities such as the magnetization or the
internal energy can be obtained from the 
corresponding left $\left<\psi^L\right|$
and right $\left|\psi^R\right>$ eigenvectors of the transfer matrix 
${\cal T}_M$.\cite{wang}

Let us emphasize that the method provides results free of finite
size errors, the thermodynamic limit being obtained automatically, due to the 
exponent $N/2$ in Eq. (\ref{partfct}). Remaining errors are the 
systematic ${\cal O}(\epsilon^2)$ Trotter error and the systematic error introduced
by the truncation of the basis set for the density matrix.\cite{white} 

\subsubsection{Imaginary time correlation function}
We now turn to the imaginary time correlation function (\ref{itgf}) and
restrict ourselves to $G_{ij}^z(\tau)$, the extension to $G_{ij}^x(\tau)$
being straightforward. For convenience, we express (\ref{itgf}) as
\begin{eqnarray}
G_{ij}^z (\tau)&=&\frac{{\cal N}_{ij}(\tau)}{Z}\nonumber\\
{\cal N}_{ij}(\tau)&=&{\rm Tr}\left(S_i^z e^{-\tau H}
S_j^z e^{-(\beta-\tau)H}\right)
={\rm Tr}\left(S_i^z (e^{-\epsilon H})^{k}
S_j^z (e^{-\epsilon H})^{M-k}\right),\label{decomp_a}
\end{eqnarray}
where $\tau=\epsilon k$ and $k=0,1,\ldots,M-1 $. Using the 
following symmetric decomposition:
\begin{equation}
e^{-\epsilon H}=e^{-\frac \epsilon 2H_o}e^{-\epsilon H_e}
e^{-\frac \epsilon 2H_o}+{\cal O} (\epsilon^2),\label{decomp_b}
\end{equation}
we have 
\begin{eqnarray}
{\cal N}_{ij}(\tau)
&= {\rm Tr}&\,(\,
\underbrace{e^{-\frac \epsilon 2H_o}S_i^z e^{-\frac \epsilon 2H_o} e^{-\epsilon H_e}}_1
\underbrace{e^{-\epsilon H_o}e^{-\epsilon H_e}}_{2} \cdots
\underbrace{e^{-\epsilon H_o}e^{-\epsilon H_e}}_{k}
\nonumber\\
&&
\underbrace{e^{-\frac \epsilon 2H_o} S_j^z e^{-\frac \epsilon 2H_o}
e^{-\epsilon H_e}}_{k+1}
\underbrace{e^{-\epsilon H_o}e^{-\epsilon H_e}}_{k+2} \cdots
\underbrace{e^{-\epsilon H_o}e^{-\epsilon H_e}}_M\,)\,.
\end{eqnarray}
up to ${\cal O} (\epsilon^2)$ corrections.

As for the partition function, we insert identities along
the horizontal Trotter slices, permute the summation indices 
in the trace and form local transfer matrix chains along the
Trotter direction to obtain an expression similar to Eq.(\ref{partfct}):
\begin{equation}
{\cal N}_{ij}(\tau)={\rm Tr}
[{\cal T}_M^{ij}(k)~({\cal T}_M)^{(\frac N 2-\left[\frac j 2\right]
+\left[ \frac i 2\right]-1)}] \label{Nij2}.
\end{equation}
Here $[\frac i 2]$ is the closest integer larger than or equal to $\frac i 2$.
In defining ${\cal T}_M^{ij}(k)$, we have to differentiate the case 
$j=i,i+1$ from $j>i+1$. First, however, we need to define the local spin 
transfer matrices
\begin{eqnarray}
&&\tau^z_{l} (\sigma^{i}_{k},\sigma^{i}_{k+1}| \sigma^{i+1}_{k},
\sigma^{i+1}_{k+1})=
\langle s_{k}^{i},s_{k}^{i+1}|{\rm e}^{-\frac \epsilon 2 h_{i}}
S_{l}^z{\rm e}^{-\frac \epsilon 2h_{i}}|
s_{k+1}^{i},s_{k+1}^{i+1}\rangle\label{trans_o}
\label{locspintm}\\
&&\tau^z_{lm} (\sigma^{i}_{k},\sigma^{i}_{k+1}| \sigma^{i+1}_{k},
\sigma^{i+1}_{k+1})=
\langle s_{k}^{i},s_{k}^{i+1}|{\rm e}^{-\frac \epsilon 2 h_{i}}
S_{l}^z S_{m}^z{\rm e}^{-\frac \epsilon 2 h_{i}}| s_{k+1}^{i},
s_{k+1}^{i+1}\rangle
\label{loc2spintm}
\end{eqnarray}
where $l,m=i,i+1$. Then, for $j>i+1$,
\begin{equation}
{\cal T}_M^{ij}(k)={\cal T}_M^{i}(0)~({\cal T}_M)^
{\left[\frac j 2\right]-\left[\frac i 2\right]-1} ~{\cal T}_M^{j}(k)
\label{Tmijk1}
\end{equation}
with
\begin{eqnarray}
&&\langle \sigma^{i}_1\cdots\sigma^{i}_{2M}| {\cal T}^{l}_M(k)|\sigma_1^{i+2}
\cdots\sigma^{i+2}_{2M} \rangle \label{Tmik}\\
&=&\sum_{\{\sigma^{i+1}_n\}} \tau_{l} (\sigma^i_{2k+1},
\sigma^i_{2k+2}|\sigma^{i+1}_{2k+1},
\sigma^{i+1}_{2k+2})\tau(\sigma^{i+1}_{2k+2},
\sigma^{i+1}_{2k+3} |\sigma^{i+2}_{2k+2},\sigma^{i+2}_{2k+3})
\label{transfs}\nonumber\\
&\times&\prod_{k'\in[1,M]}^{k'\neq k+1}
\tau (\sigma^i_{2k'-1},\sigma^i_{2k'}|\sigma^{i+1}_{2k'-1},
\sigma^{i+1}_{2k'})
\tau(\sigma^{i+1}_{2k'},\sigma^{i+1}_{2k'+1} |\sigma^{i+2}_{2k'},
\sigma^{i+2}_{2k'+1}).
\nonumber
\end{eqnarray}
Again, these formulas can be nicely understood
graphically as in Figs. \ref{checkb}(c) and \ref{checkb}(d), 
corresponding to Eqs.
(\ref{Nij2})-(\ref{Tmik}) for the case $N=6$, $M=3$ ,$i=1$ and $j=3,4$.
In this case, the two local spin transfer matrices sit on different
transfer matrix chains.

On the contrary, for $j=i,i+1$, they belong
to the same chain:
\begin{eqnarray}
&&\langle \sigma^{i}_1\cdots\sigma^{i}_{2M}| {\cal T}_M^{ij}(k)|\sigma_1^{i+2}
\cdots\sigma^{i+2}_{2M} \rangle \label{TijMdef2}\\
&=&\sum_{\{\sigma^{i+1}_n\}} \tau^z_{l} (\sigma^{i}_{1},
\sigma^{i}_{2} |\sigma^{i+1}_{1},  \sigma^{i+1}_{2})
\tau(\sigma^{i+1}_{2},\sigma^{i+1}_{3}|\sigma^{i+2}_{2},
\sigma^{i+2}_{3})\nonumber\\
&\times&\tau^z_{m} (\sigma^i_{2k+1},\sigma^i_{2k+2}
|\sigma^{i+1}_{2k+1},\sigma^{i+1}_{2k+2})
\tau(\sigma^{i+1}_{2k+2},\sigma^{i+1}_{2k+3}
|\sigma^{i+2}_{2k+2},\sigma^{i+2}_{2k+3})\nonumber\\
&\times&\prod_{k'\in[1,M]}^{k'\neq 1,k+1}
\tau(\sigma^i_{2k'-1},  \sigma^i_{2k'}|\sigma^{i+1}_{2k'-1},\sigma^{i+1}_{2k'})
\tau(\sigma^{i+1}_{2k'},\sigma^{i+1}_{2k'+1}|\sigma^{i+2}_{2k'},
\sigma^{i+2}_{2k'+1})
\nonumber
\end{eqnarray}
when $k=1,2,\ldots,M-1$ and
\begin{eqnarray}
&&\langle \sigma^{i}_1\cdots\sigma^{i}_{2M}| 
{\cal T}_M^{ij}(k)|\sigma_1^{i+2}\cdots\sigma^{i+2}_{2M} \rangle \\
&=&\sum_{\{\sigma^{i+1}_n\}} \tau^z_{lm} (\sigma^{i}_{1},  
\sigma^{i}_{2} |\sigma^{i+1}_{1},  \sigma^{i+1}_{2})
\tau(\sigma^{i+1}_{2},\sigma^{i+1}_{3}|\sigma^{i+2}_{2},
\sigma^{i+2}_{3})\nonumber\\
&\times&\prod_{k'\in[1,M]}^{k'\neq 1}
\tau(\sigma^i_{2k'-1},\sigma^i_{2k'}|\sigma^{i+1}_{2k'-1},\sigma^{i+1}_{2k'})
\tau(\sigma^{i+1}_{2k'},\sigma^{i+1}_{2k'+1}|\sigma^{i+2}_{2k'},
\sigma^{i+2}_{2k'+1}),\nonumber
\end{eqnarray}
for $k=0$, corresponding to a static correlation.

Finally, we obtain for the imaginary time correlation
\begin{equation}
G_{ij}^z(\tau)= \frac {{\rm Tr}[{\cal T}_M^{ij}(k)
~({\cal T}_M)^{(\frac N 2-\left[\frac j 2\right ]+\left[\frac i 2\right]-1)}]}
{{\rm Tr}({\cal T}_M)^{\frac N 2}}
\end{equation}
In the limits $N\rightarrow\infty$ and $|i-j|\ll N$, this reduces to
\begin{equation}
G_{ij}^z(\tau)= \frac {\langle\psi^L|{\cal T}_M^{ij}(k)|\psi^R\rangle}
{\lambda_{\rm max}^{\left[\frac j 2\right]-\left[\frac i 2\right]+1}}.
\label{Gijg}
\end{equation}
For systems with finite correlation length $\xi$,
$G^{z}_{ij}(\tau)\sim \left< S^z_i \right> \left<S^z_j\right>$ when
$|j-i|\gg\xi$, which can be verified systematically as
$\xi$ is determined from 
$\xi^{-1}= \frac 1 2 \ln\left|\frac{\lambda_n}{\lambda_{\rm max}}\right|$. 
$\lambda_n$ is the next largest eigenvalue of ${\cal T}_M$.

\subsubsection{Renormalization of transfer matrices}

\begin{figure}
\epsfxsize=12cm
\epsffile{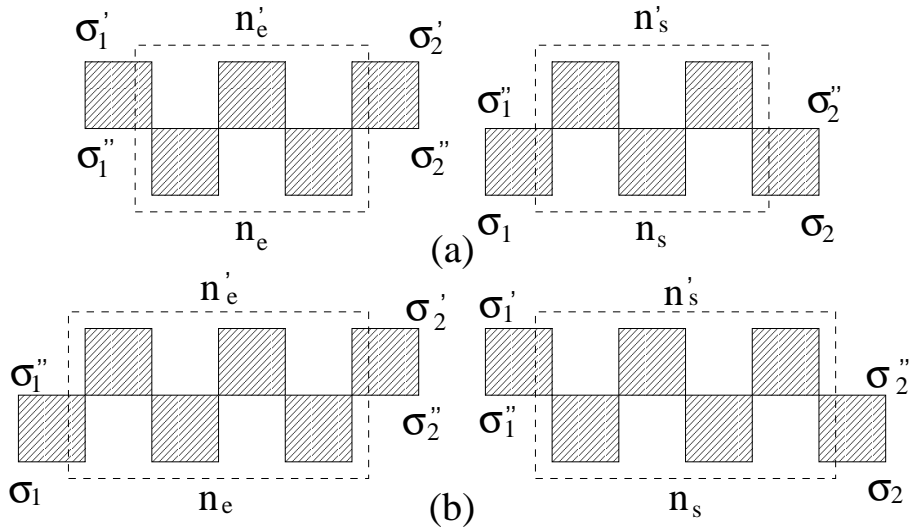}
\caption{The superblock ${\cal T}_M$ is cut in a system ${\cal S}$ and
environment ${\cal E}$ block for M=5 odd (a), M=6 even (b).}
\label{halves}
\end{figure}

Now that we have defined the relevant transfer matrices, we
must explain how to construct them in practise.
The purpose is to add
successively new $\tau$ plaquettes to grow ${\cal T}_M$
in the Trotter direction (in the same spirit real sites are added 
to the Hamiltonian in the standard $T=0$ DMRG algorithm\cite{white}) and
simultaneously find a renormalization procedure that keeps the
dimension of the matrix ${\cal T}_M$ fixed as the temperature
is lowered.
For this purpose, we cut ${\cal T}_M$ schematically in two halves
as shown in Fig. \ref{halves} for the two generic cases of $M$ 
odd (a) and even (b). In the DMRG language, ${\cal T}_M$ is called 
the superblock, the right inner block (dashed line in Fig.\ref{halves})
plus the left edge $(\sigma_1,\sigma_1^{''})$ the system 
and the left inner block plus the right edge 
$(\sigma_2,\sigma_2^{''})$ the environment. We use $n_s$ and 
$n_e$ to label the basis sets in the inner
of the system and environment blocks, respectively. The states at the
left and right edges are labelled by $\sigma_1$ and $\sigma_2$. With
this notation,
we denote the elements of the right transfer matrix 
by ${\cal S}_o(\sigma_1'',n_s',\sigma_2'';\sigma_1,n_s,\sigma_2)$ or
${\cal S}_e(\sigma_1',n_s',\sigma_2'';\sigma_1'',n_s,\sigma_2)$ depending
on whether the system consist of an even ($e$) or odd ($o$)
number of $\tau$ plaquettes. 
When adding a new plaquette, the elements of the new system
are given by the following recursion relation:
\begin{eqnarray}
&&{\cal S}_e(\sigma_1',\tilde n_s',\sigma_2'';\sigma_1'',\tilde n_s,\sigma_2)=
\sum_{\sigma''} \tau(\sigma_1',\sigma'|\sigma_1'',\sigma'')
{\cal S}_o(\sigma'',n_s',\sigma_2'';\sigma,n_s,\sigma_2),
\\
&&{\cal S}_o(\sigma_1'',\tilde n_s',\sigma_2'';\sigma_1,\tilde n_s,\sigma_2)
=\sum_{\sigma''}
\tau(\sigma_1'',\sigma''|\sigma_1,\sigma){\cal S}_e(\sigma',n_s',
\sigma_2'';\sigma'',n_s,\sigma_2),
\label{system}
\end{eqnarray}
where $\{ \left|\tilde n_s\right> \}= \{\left|\sigma\right> \}\otimes
\{ \left|n_s \right> \}$. Initially, when $M=2$,
${\cal S}_e(\sigma_1',\sigma',\sigma_2'';\sigma_1'',\sigma,\sigma_2)=
\sum_{\sigma''}\tau(\sigma_1',\sigma'|\sigma_1'',\sigma'')
\tau(\sigma'',\sigma_2''|\sigma,\sigma_2)$.

Similarly, the elements of the left transfer 
matrix
${\cal E}_o(\sigma_1',n_e',\sigma_2';\sigma_1'',  n_e,\sigma_2'')$ or
${\cal E}_e(\sigma_1'', n_e',\sigma_2';\sigma_1,n_e,\sigma_2'')$,
are given by
\begin{eqnarray}
&&{\cal E}_e(\sigma_1'',\tilde n_e',\sigma_2';\sigma_1,\tilde n_e,\sigma_2'')=
\sum_{\sigma''} \tau(\sigma_1'',\sigma''|\sigma_1,\sigma)
{\cal E}_o(\sigma',n_e',\sigma_2';\sigma'',n_e,\sigma_2''),
\\
&&{\cal E}_o(\sigma_1',\tilde n_e',\sigma_2';\sigma_1'',\tilde n_e,\sigma_2'')
=\sum_{\sigma''}
\tau(\sigma_1',\sigma'|\sigma_1'',\sigma'')
{\cal E}_e(\sigma'',n_e', \sigma_2';\sigma,n_e,\sigma_2'')
\label{environment}
\end{eqnarray}
after one plaquette has been added, $\{\left|\tilde n_e\right>\}
=\{\left|\sigma\right>\}\otimes \{\left|n_e\right>\}$.

When the number of states in $\{\left|\tilde n_s\right>\}$ 
($\{\left|\tilde n_e\right>\}$) exceeds a given number $m$,
the transfer matrices ${\cal S}_{e,o}$ (${\cal E}_{e,o}$) are renormalized by:
\begin{eqnarray}
&&{\cal A}_e(\sigma_1',n_s',\sigma_2'';\sigma_1'',n_s,\sigma_2)
=\sum_{\tilde n_s'\tilde n_s}
O_{\cal A}^{l}(n_s',\tilde n_s'){\cal A}_e(\sigma_1',\tilde n_s',\sigma_2''
;\sigma_1'',\tilde n_s,\sigma_2) O_{\cal A}^r(\tilde n_s,n_s),
\\
&&{\cal A}_o(\sigma_1'',n_s',\sigma_2'';\sigma_1,n_s,\sigma_2)
=\sum_{\tilde n_s'\tilde n_s}
O_{\cal A}^{l}(n_s',\tilde n_s'){\cal A}_o(\sigma_1'',\tilde n_s',\sigma_2''
;\sigma_1,\tilde n_s,\sigma_2)
O_{\cal A}^r(\tilde n_s,n_s),
\label{renorm}
\end{eqnarray}
${\cal A}={\cal S}$ for the system block or ${\cal E}$ for the environment
block, $n_s',n_s=1,2,\ldots,m$.
The transformation matrices $O^{l}_{\cal A}$ and  $O^{r}_{\cal A}$
are constructed from the $m$ largest eigenvectors of the corresponding reduced,
non-symmetric density
matrices given in terms of the left and right eigenvectors of ${\cal T}_M$
by:
\begin{equation}
\rho_{s}= {\rm Tr}_{n_e,\sigma_2} |\psi^R\rangle \langle \psi^L|,~~~
\rho_{e}= {\rm Tr}_{n_s,\sigma_1} |\psi^R\rangle \langle \psi^L|,
\end{equation}
for the system and the environment. In terms of ${\cal S}$ and ${\cal E}$, the
renormalized transfer matrix 
${\cal T}_M$ corresponding to the superblock (Fig. \ref{halves}) is given by
\begin{eqnarray}
{\cal T}_M(n_e',\sigma_2',\sigma_1',n_s';
n_e,\sigma_2,\sigma_1,n_s)=\left\{
\begin{array}{rl}\displaystyle{\sum_{\sigma_1'',\sigma_2''}}
 &{\cal E}_o(\sigma_1',n_e',\sigma_2';\sigma_1'',n_e,\sigma_2'')\\
\times&{\cal S}_o(\sigma_1'',n_s',\sigma_2'';\sigma_1,n_s,\sigma_2),\\
\displaystyle{\sum_{\sigma_1'',\sigma_2''}}
 &{\cal E}_e(\sigma_2'',n_e',\sigma_2';\sigma_1,n_e,\sigma_2'')\\
\times&{\cal S}_e(\sigma_1',n_s',\sigma_2''; \sigma_1'',n_s,\sigma_2)
\end{array}\right.\label{superblock}
\end{eqnarray}
for $M/2$ odd or even.

Many systems of interest have spatial reflection symmetry.
For these, we can obtain the left transfer matrix from transposition
of the right one so that for instance, ${\cal E}_o(\sigma_1',n_e',\sigma_2';
\sigma_1'',n_e,\sigma_2'')
={\cal S}_o(\sigma_1'',n_e,\sigma_2'';\sigma_1,n_e',\sigma_2)$. Consequently, 
the left eigenvector of $\langle \psi^L|$ can be obtained from
$|\psi^R\rangle$:  $\psi^L(n_s,\sigma_2,
\sigma_1,n_e) = \psi^R(n_e,\sigma_2,\sigma_1, n_s)$.
For systems without reflection symmetry such as zigzag or
dimerized chains, left and right transfer matrices and eigenvectors
must be evaluated separately. 

\begin{figure}
\epsfxsize=12cm
\epsffile{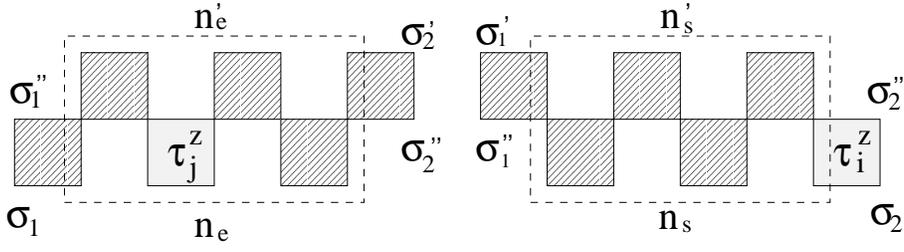}
\caption{In the case $j=i$ or $i+1$, ${\cal T}_M^{ij}(k=2)$ is cut 
in ${\cal S}_e^i$ and
${\cal E}_e^j(k)$ for $M=6$, $i=1$.}
\label{halvsp}
\end{figure}

As a last technical step, we have to explain how to renormalize 
${\cal T}_M^{ij}(k)$, 
needed for evaluating $G^z_{ij}(\tau)$. Again, we must distinguish two cases:

For $j=i,i+1$, the two local spin transfer matrices are located on the
same transfer matrix chain. In this case, we can separate ${\cal T}_M^{ij}(k)$
into left ${\cal E}_{e,o}^{j}(k)$ and right ${\cal S}_{e,o}^{i}$ parts, 
similar to what we did for ${\cal T}_M$. 
As $G^z_{ij}(\tau)=G^z_{ij}(\beta-\tau)$,
it is sufficient if $k=0,1,\ldots,M/2$ so that there is always exactly one
local spin transfer matrix in each half of ${\cal T}_M^{ij}(k)$ 
(Fig. \ref{halvsp}.)
The blocks ${\cal E}_{e,o}^{j}(k)$ and ${\cal S}_{e,o}^{i}$ are renormalized
according to Eqs. (\ref{system})-(\ref{renorm}) except that 
$\tau$ has to be substituted with $\tau^z_{l}$ at the 
appropriate steps. The static, $k=0$, case has to be treated separately. 

\begin{figure}
\epsfxsize=12cm
\epsffile{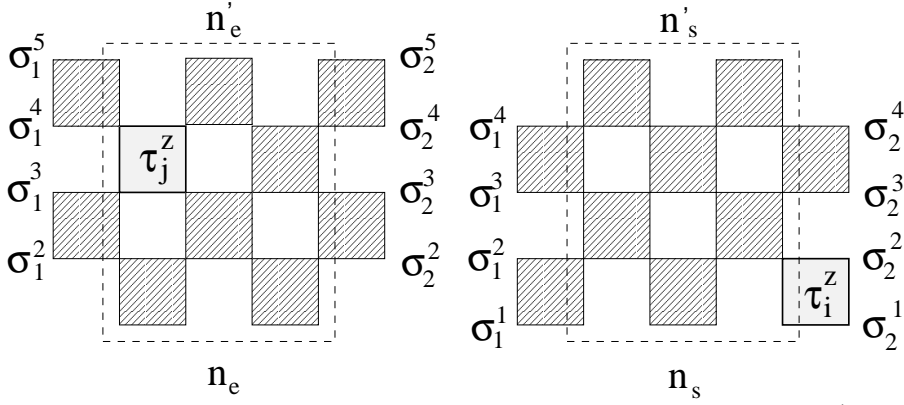}
\caption{When $j.i+1$, additional internal $\sigma$ variables are 
needed in both ${\cal S}_e^i$ and ${\cal E}_e^j(k)$. Here, $i=1$, $j=3$ or $4$,
$k=2$ and $M=6$.}

\label{halvsp2}
\end{figure}

When $j>i+1$, ${\cal T}_M^{ij}(k)$ consists of 
$(\left[\frac j 2\right]$-$\left[\frac i 2\right]$+$1)$ adjacent 
parallel chains
which must be renormalized as a whole. Again, they can be cut into halves
${\cal E}_{e,o}^{j}(k)$ and  ${\cal S}_{e,o}^{i}$, with 
$n_i=2(\left[\frac j 2\right]$-$\left[\frac i 2\right])+1$ internal real 
space lines instead of one in the previous case (Fig. \ref{halvsp2}). 
These can be renormalized
according to Eqs. (\ref{system})-(\ref{renorm}) with the 
proper modifications. However, the storage of ${\cal E}_{e,o}^{j}(k)$,
${\cal S}_{e,o}^{i}$ will be increased by a factor $(2S+1)^{2(n_i-2)}$ which will 
eventually become restrictive for large $j-i$. If this is the case, we
approximate $G_{ij}^z(\tau)$ by substituting the renormalized ${\cal T}_M$,
${\cal T}_M^{i}(0)$
and ${\cal T}_M^{j}(k)$ in the expression for ${\cal T}_M^{ij}(k)$ (Eq.(\ref{Tmijk1}))
and multiply them successively to $\left|\psi^R\right>$ in Eq. (\ref{Gijg}).
The accuracy of the obtained result can be controlled by varying
the number of states m.

For a realistic description of NMR experiments, which involves at most $j=i,i+1,i+2$
correlations, the calculation can be easily 
performed for systems having up to three states per site.

\subsection{Analytical continuation}
The analytical continuation is concerned with the inversion of the integral 
equation (\ref{intequ0}) whose
principal feature is that the kernel $K(\tau,\omega)$ is very
singular. This renders a numerical inversion particularly sensible to 
errors in $G^{\alpha}_{ij}(\tau)$ which are exponentially amplified in 
the result. 
Before discussing the probabilistic approach to ill-posed
inversion problems we will briefly discuss the SVD method because
it gives insight on how ill-posed the problem is.

Let us first discretize the $\tau$ and $\omega$ variable and restate 
Eq. (\ref{intequ0}) in matrix
form, omitting the superfluous subscripts for now,
\begin{equation}
G(\tau_i)=\sum_j K_{ij} S(\omega_j),~~~i=1,...,N,~~~j=1,...,M
\label{intequ1}
\end{equation}
At this stage we should mention that it is straightforward to
incorporate knowledge on the  
derivatives of $G^{(n)}(\tau)=\frac{d^n}{d\tau^n}G(\tau)$ by just adding
more equations
\begin{eqnarray}
G^{(n)}(\tau_l)=\sum_j K^{(n)}_{lj} S(\omega_j),~~~l=1,...,N',~~~j=1,...,M
\end{eqnarray}
to the linear system (\ref{intequ1}). The formal problem is unchanged,
only the vector of data points and the kernel are larger in the first
index.

The SVD decomposition of the matrix $K$ is $K_{ij} = \sum_{l=1}^M
U_{il} \Lambda_l V_{jl}$ where $V$ is an $M\times M$ orthogonal
matrix, while the $N\times M$ matrix $U$ is merely
column-orthonormal, i.e. $U^T U$ is the $M\times M$ unit matrix.
The $\Lambda_l^2$ correspond to the $M$ eigenvalues of the matrix
$K^T K$. Formally, the solution of equation (\ref{intequ1}) is:
\begin{equation}
S(\omega_j)=\sum_{l=1}^M \sum_{i=1}^N V_{jl} \frac{1}{\Lambda_l}
U_{il} G(\tau_i)\label{sumsvd}
\end{equation}
One immediately anticipates the catastrophe when the eigenvalues
$\Lambda_l$ become small and $G(\tau)$ contains errors. Since
$K_{ij}\ge0$, small eigenvalues correspond to rapidly oscillating
eigenvectors of $K^T K$ which therefore couple strongly to the
noise. To illustrate the ill-posed nature of the analytic
continuations, consider the situation where $N=M=100$. The
condition number , i.e. the ratio of the largest to the smallest
$|\Lambda_l|$, is greater than $10^{17}$ for the entire range of
interesting $\beta$-values . In other words, even the errors introduced
by the finite machine-accuracy are sufficient to make the direct inversion
useless.
For $\beta=1$ only $5\%$ of the
eigenvalues are greater than $10^{-8}$ and for $\beta=10$ the
percentage is $10\%$. The situation is roughly unchanged when
including derivatives. The straightforward application of
Eq. (\ref{sumsvd}) yields results which are orders of magnitude
too large. The natural way to regularize the sum (Eq. \ref{sumsvd}) is
by truncating it at $l_{cut}$ so that the error $\delta(\sum_i
G(\tau_i)U_{il_{cut}})\ll \sum_i G(\tau_i)U_{il_{cut}}$. Typical
values of $l_{cut}$ for the transfer matrix DMRG data are 7 to 10. 
The drawback of
this ad-hoc truncation-scheme is that a major part of the vector
space for $S(\omega_j)$ is lost.
A further disadvantage of the SVD-approach is that it does not
enforce positivity of the solutions.
It should be noted that the
SVD-approach is equivalent to Tichonov-regularization, where the
$L^2$-norm of the image is the regularization criterion. We should
also mention that we tested the Pade approximation. It appeared
that it did well in some situations (isotropic case), but it
failed in others (gaped phase). Therefore, we will not
discuss this method further.

There is a wide-spread misconception about ill-posed inversion
problems, or inductive inference problems in general. It makes no
sense to ask for the true function $f(\omega)$. There is no
chance, whatsoever, to infer the true result! All inference
schemes yield results which can in principle deviate widely from
the unknown true result. The correct question to be asked and
which can uniquely be answered is rather: what is the distribution
of functions $f(\omega)$ compatible with the noisy and incomplete
data and all our prior knowledge. In other words, we should aim
for the probability density $p(f(\omega)|D,I)$ for a function
$f(\omega)$ in the light of the transfer matrix DMRG data $D$ 
and additional
prior-knowledge $I$, such as sum-rules and positivity constraints.
The elementary product-rule of probability theory allows us to
determine this probability consistently:
\begin{equation}
p(f(\omega)|D,I) = p(f(\omega)|I) p(D|f(\omega),I) /p(D|I)
\label{posterior}
\end{equation}
in terms of the likelihood $p(D|f(\omega),I)$, the prior
$p(f(\omega)|I)$ and the normalization $p(D|I)$. The likelihood
stands for the probability for the data $D$, assuming that
$f(\omega)$ is the exact function. The likelihood deviates from a
delta-functional if the data suffer from statistical noise or
unknown systematic errors, like in the present case. For the
likelihood the source of the missing information does not matter.
As long as we know nothing about the features of the systematic
errors, the data have to be considered as the mean, and the errors
as the variance of the likelihood-distribution function
\cite{jaynes}. Hence, like in the case of uncorrelated
normal-distributed data, the likelihood reads
$p(D|f(\omega),I)\propto \exp(-\chi^2 / 2)$, with
\begin{equation}
\chi^2=\sum_i\frac{|G(\tau_i)-\sum_j
K_{ij}S(\omega_j)|^2}{\sigma_i^2}\qquad.
\end{equation}
The prior distribution $p(f(\omega)|I)$ quantifies our knowledge
about the solution $f$ prior to our knowing the data $D$. For a
general scheme, the only reliable prior knowledge is that $f$ is a
positive additive distribution function for which the adequate
probability distribution is the entropic prior
$p(f(\omega))\propto \exp(\alpha S)$,\cite{skilling} with $S$
being the information divergence or relative entropy
\begin{equation}
S=\sum_j \left(f(\omega_j)-m(\omega_j)-f(\omega_j) \ln \left(
\frac{f(\omega_j)}{m(\omega_j)} \right) \right)K_{0j}
\label{entropy}
\end{equation}
of the function $f(\omega_j)$ relative to a default model
$m(\omega_j)$.
The factor $K_{0j}$
accounts for the negative frequency contribution consistently with
the detailed balance condition. Maximizing the posterior
probability $p(f(\omega)|D,I)$ is equivalent to maximizing the
functional
\begin{equation}
\phi=\alpha S-\frac{1}{2} \chi^2
\end{equation}
When maximizing $\phi$, $S$ essentially regularizes the solution
such that: (i) it stays positive and, (ii) structure relative to
the default model is penalized, depending on the parameter
$\alpha$. MaxEnt yields the most uncommittal solution which shows
only structure if it is
significantly supported by the data.
In ''classic" MaxEnt, $\alpha$ is determined self-consistently
using the rules of probability theory, i.e.
such that the solution $f(\omega_j)$ is the most
probable in the light of the input data (details can be found in
Ref.[\CITE{wvl,gub}]). We have only considered a flat
$m(\omega_j)$ for $\omega_j>0$.

As mentioned above, in the transfer matrix calculation 
(as opposed to QMC), the
errors $\sigma_i$ are systematic but unknown. They are due to the
truncation of the basis set and to a finite Trotter step. Since
the values for $\sigma_i$ are not known we have to determine the
probability $p(\sigma_i|D, I)$ for $\sigma_i$ using the rules of
probability theory\cite{bretthorst}. Typically, $\sigma_i \sim
10^{-6}$ for $\beta=16$. So doing, the reconstructed imaginary
time correlation agrees with the DMRG data up to $\sim 10^{-6}$ or
better.

\section{Results} 
\subsection{XY model: Longitudinal autocorrelation}
\label{sxyzz}

The XY model ($\Delta=0$), which can be mapped onto a free fermion model via a 
Jordan- Wigner transformation, is useful for tests because its longitudinal
$zz-$correlations in $(q,\omega)$ can be expressed in closed form
\cite{niekat} at any temperature $T$.

The corresponding imaginary time autocorrelation function can be 
represented as 
\begin{equation}
G_{ii}^{z}(\tau)=\left(\frac{1}{\pi} \int_0^{\pi}
\frac{e^{\tau \cos q}}
{1+e^{\beta \cos q}}dq\right)^2
\end{equation}

\begin{figure}
\epsfxsize=12cm
\epsffile{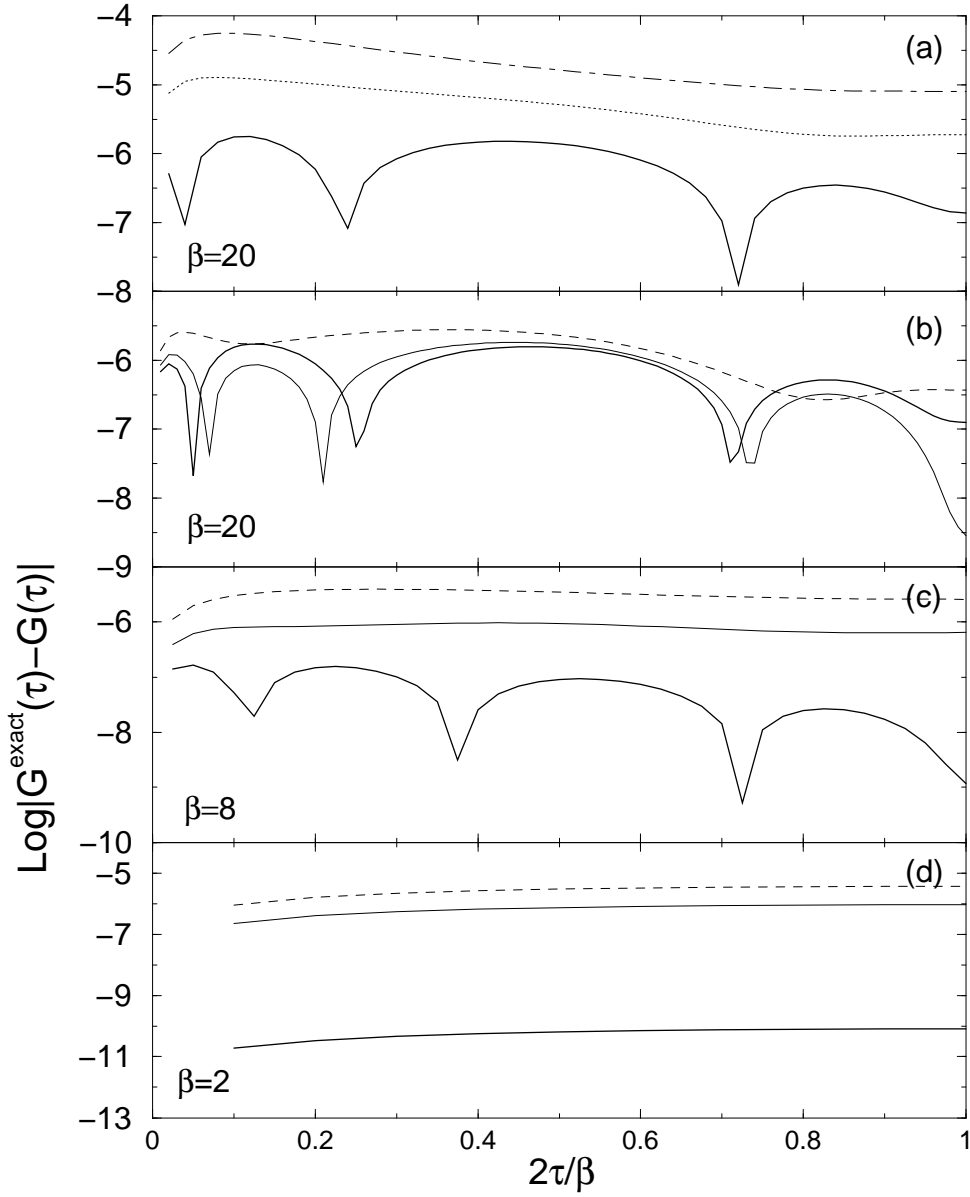}
\caption{Exact $G_{ii}(\tau)$ versus transfer matrix result with $m=100$ states. 
The thick solid line 
always represents the result extrapolated from the two finite 
$\epsilon$ values. In (a), $\epsilon=0.2$(dot-dashed), $0.1$(dotted) 
and in (b)-(d) $\epsilon=0.025$(dashed), $0.05$(thin solid).}
\label{gtexvsdmrg}
\end{figure}

In Fig. \ref{gtexvsdmrg}, we compare this exact result with 
the transfer matrix DMRG data for
$\beta=2,8,20$ on a logarithmic scale. The reversed peaks are
merely artifacts due to the change of sign in the argument of the logarithm. 
For all the calculations, we have kept $m=100$ states 
in the density matrix so that the truncation error
($1-\sum_m\rho_m$) is smaller than $10^{-7}$ for the largest
Trotter number M=800 ($\beta=20$ when $\epsilon=0.025$).
In order to reduce the Trotter error, we have done a linear 
$\epsilon^2 \rightarrow 0$ extrapolation 
(this requires commensurate values of $\epsilon$).
This procedure is justified as long as the systematic errors induced by the
truncation are negligible compared to the Trotter errors. This is typically 
the case at high temperatures (small truncation error,
Fig. \ref{gtexvsdmrg}(d))
or when $\epsilon$ is large (large Trotter error, Fig. \ref{gtexvsdmrg}(a)).
Fig. \ref{gtexvsdmrg}(b) is an example where the $\epsilon^2$ 
extrapolation fails
because the systematic errors become comparable to the Trotter error itself
($M=800$ Trotter steps are needed to reach
$\beta=20$ when $\epsilon=0.025$). In such a case, nothing can 
be gained from the extrapolation and it is justified to use the 
smallest $\epsilon$ data for the analytical continuation.
Notice that the result in Fig. \ref{gtexvsdmrg}(a) obtained from the 
$\epsilon=0.1,0.2$ data is as precise as the $\epsilon=0.025$ calculation.
This can be exploited for models with more degrees of freedom per site, 
where small $\epsilon$ calculations cannot be afforded and values of 
$\epsilon=0.1$ or larger may be needed to reach the low temperature regime.

Before doing the continuation to real frequencies, we
should warn that because the XY model essentially describes free fermions, 
its $zz-$correlation function 
is not generic of a true interacting system and exhibits some peculiar behavior.
For instance, the $zz-$autocorrelation has a sharp steplike cutoff 
at $\omega=2$ and a $\omega=0$ logarithmic divergence for $\beta=0$.
As we are not interested in reproducing step discontinuities,
not expected in interacting systems at finite temperature,
we artificially cutoff the integral at $\omega=2$ in Eq. (\ref{intequ0}) 
to investigate how the smooth lineshape in the interval $\omega\in [0,2]$
can be reconstructed.

\begin{figure}
\epsfxsize=12cm
\epsffile{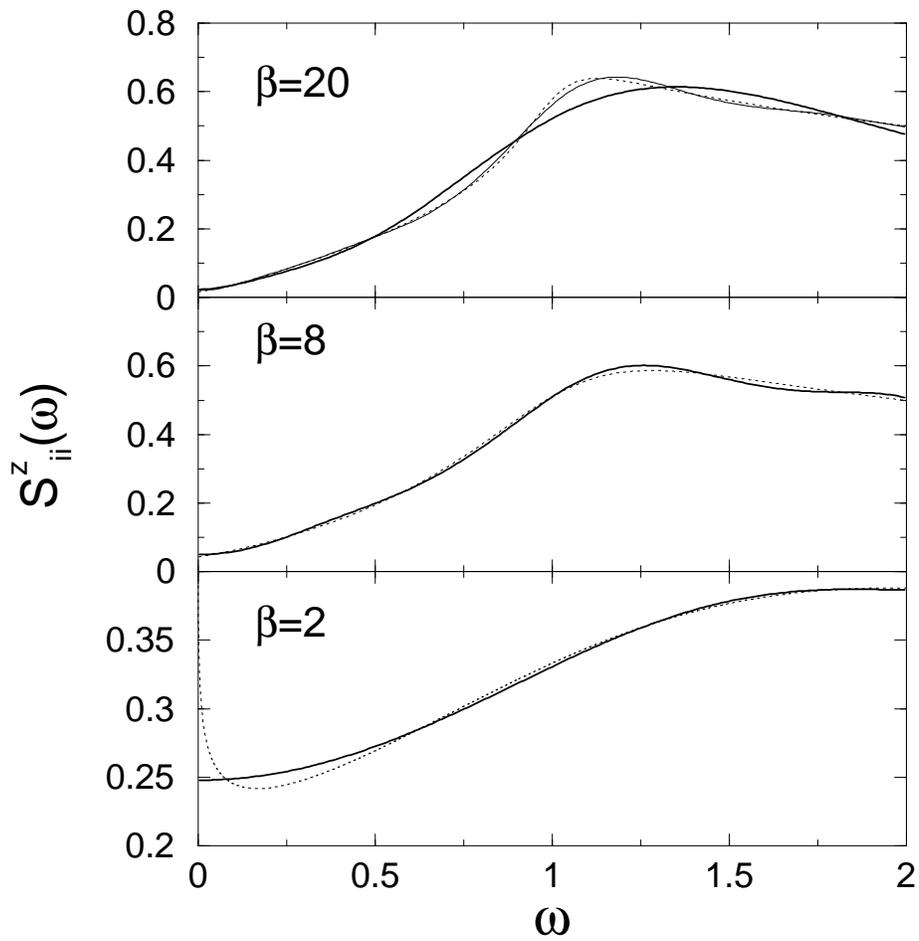}
\caption{XY model, $S^z_{ii}(\omega)$ using
MaxEnt for the analytical continuation. The dotted line is
the exact solution. The thin solid thin lines
are obtained from the exact $G^z_{ii}(\tau)$.}
\label{svdme}
\end{figure}

In Fig. \ref{svdme}, we compare the MaxEnt
results obtained from the transfer matrix DMRG data, those obtained 
from the exact $G(\tau)$ and the exact $S_{ii}^{zz}(\omega)$.
Notice that MaxEnt cannot resolve the low frequency divergence at $\beta=2$
since there is not enough spectral weight under the logarithmic divergence.
The MaxEnt results from the numerical data and those from the
exact $G(\tau)$ cannot be distinguished for $\beta=2$ and $\beta=8$.

\subsection{XY model: Transverse autocorrelation}\label{sxyxx}
After the Jordan-Wigner transformation, the $xx$-correlations become
nonlocal fermion correlations which are expressible
as Pfaffians\cite{lsm} at all temperatures. Closed form expressions 
are known only at  
$T=0,\infty$\cite{mueller}. The exact $T=0$ autocorrelations have 
singular behavior at all integer frequencies 
$\omega=l$, $l=0,1,2,...$. $S^{x}_{ii}(\omega)$ diverges as $\omega^{-1/2}$
for $\omega\rightarrow 0^+$ and as $\ln|\omega-1|$ for $\omega\rightarrow 1$.
The higher singularities are cusps, for $\omega\rightarrow 2^+$,
$S^{x}_{ii}(\omega) \sim (\omega-2)^{1/2}$. At $T=\infty$,
$S^x_{ij}(\omega)=\delta_{ij}\frac{\sqrt{\pi}}{2J} e^{-{\omega}^2/J^2}$.

\begin{figure}
\epsfxsize=14cm
\epsffile{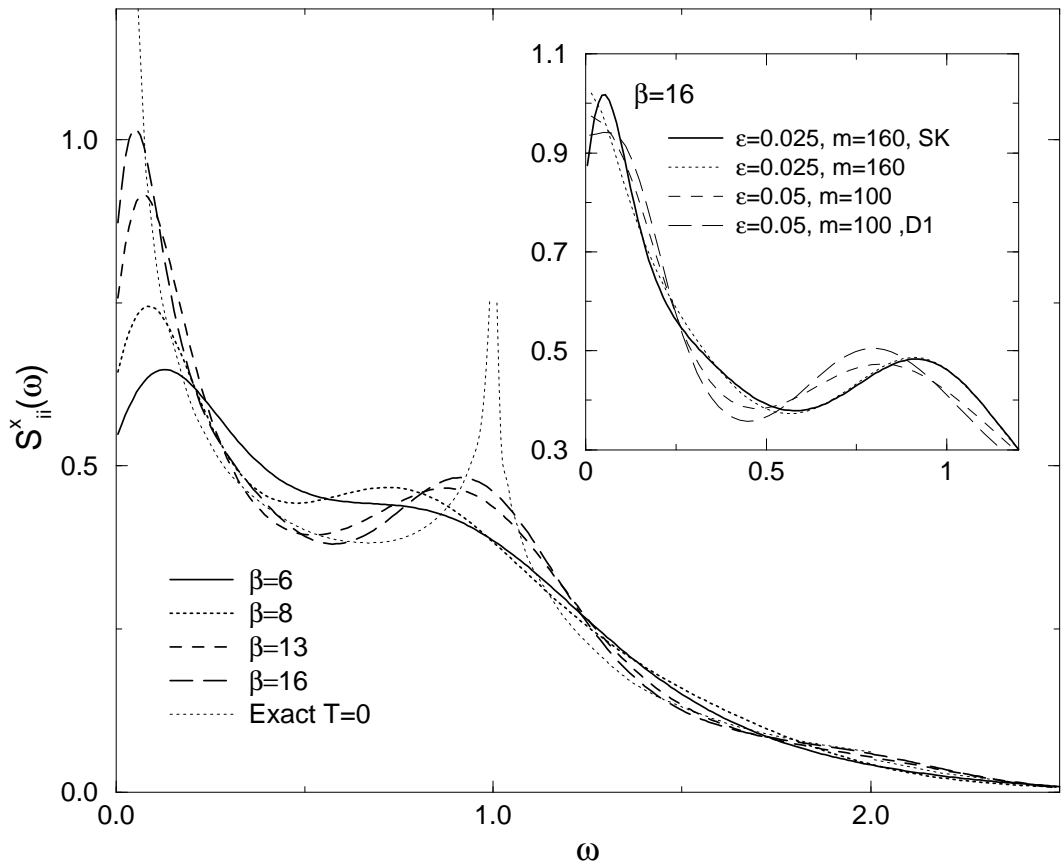}
\caption{XY model, $S^x_{ii}(\omega)$ from MaxEnt. In the transfer matrix DMRG calculations, $m=160$
states were kept and $\epsilon=0.025$. Inset: SK is gotten from the 
symmetric kernel and D1 includes $\frac{d}{d\tau}G(\tau)$.}
\label{figxx}
\end{figure}

In Fig. \ref{figxx}, we show the numerical data continued with MaxEnt 
in the intermediate to low temperature range $\beta=6-16$. 
As the temperature is lowered, one can see the emergence of two peaks, one at 
$\omega=1$ and the other near $\omega=0$. Also, we see some sign of 
the cusp at $\omega=2$. 
Reducing the temperature, the peak near $\omega=0$ continuously grows and 
shifts closer to the zero temperature singularity.
That at $\omega=1$ consistently moves towards the correct $T=0$ 
position, as long as the temperature is not too low ($\beta\leq 16$). 
When the quality of the data becomes poor, which can be due
to a large Trotter number M (low temperature), to a small basis set (small m),
or to a large Trotter step $\epsilon$, we observe a systematic shift of high frequency
structures towards lower frequencies.
This effect can be seen in the inset when only $m=100$ states are 
kept instead of $m=160$ and $\epsilon=0.05$ rather than $0.025$.

Further, we learned that it is better to reconstruct 
$S^x_{ii}(\omega)$ using the symmetric scheme.
Indeed, it is known\cite{its} that the time correlation $S^x_{ij}(t)$ decays
exponentially for all $(i,j)$ and temperatures $T$, therefore, 
$S^x_{ii}(\omega)$ is regular
at $\omega=0$. This implies that $\frac{d}{d\omega}
S^x_{ii}(0)=\frac{\beta}{2}S^x_{ii}(0)$ due to the detailed balance condition.
When reconstructing directly $S^x_{ii}(\omega)$, this relation is not fulfilled
(the two procedures are compared in the inset for $\beta=16$).

Finally, in an attempt to increase the information contained in the input 
data, we calculated $\frac{d}{d\tau}G(\tau)$ with
equal precision as $G(\tau)$. We find that the additional data does not help
much in the analytical continuation (which can be understood when looking
at the SVD decomposition of the extended kernel), except that it gives a hint
that the curve obtained by the symmetric kernel is better behaved 
near $\omega=0$. In any case, we found it more economical to improve the 
accuracy on $G(\tau)$ rather than calculate $\frac{d}{d\tau}G(\tau)$.

\subsection{Isotropic Heisenberg model}
\label{sheis}
Rigorous results for the dynamic correlations in the isotropic Heisenberg 
model ($\Delta=1$) are rare. Recently,\cite{karbach} the exact two
spinon contribution to $S^x(q,\omega)$ at $T=0$ was found. In this 
restricted subspace, the autocorrelation function has singularities 
$\ln(\omega)$ for $\omega\rightarrow 0$ and $(\ln|\pi/2-\omega|)^{3/2}$ as 
$\omega\rightarrow\pi/2$. 

\begin{figure}
\epsfxsize=12cm
\epsffile{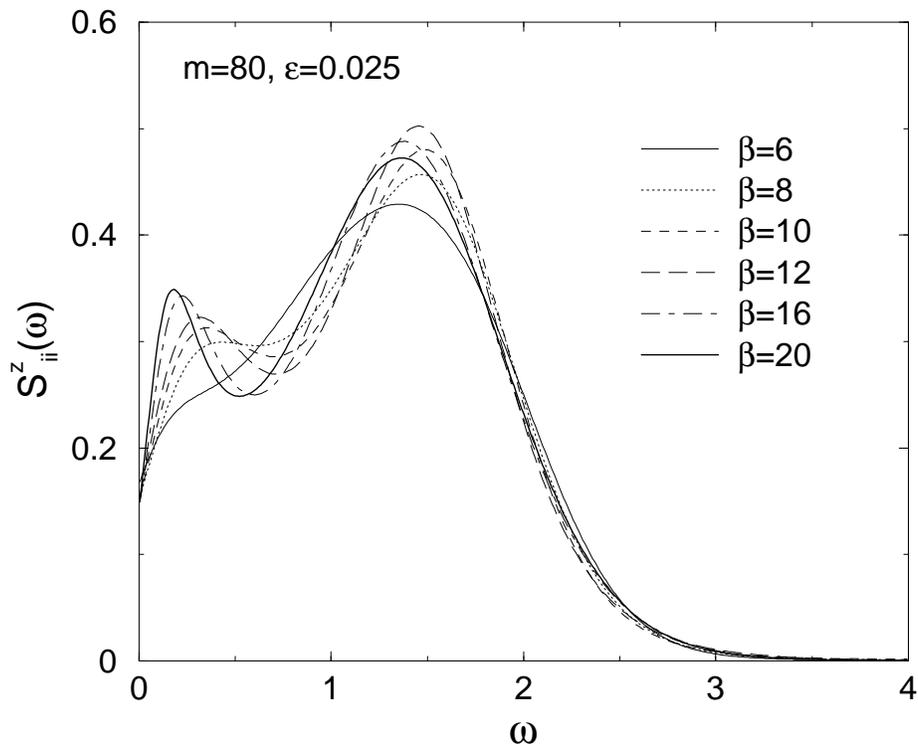}
\caption{Isotropic Heisenberg Model. MaxEnt from numerical data with $m=80$ and 
$\epsilon=0.025$.}
\label{figheis}
\end{figure}

In Fig. \ref{figheis}, we present results for $S^z_{ii}(\omega)$. 
The main peak at $\omega=\pi/2$ for $T=0$ is 
shifted to slightly lower frequencies at finite temperatures. 
As in the $xx-$correlations of the $XY-$model, 
this peak seems to move away from the exact value for the lowest 
temperatures studied, presumably due to loss of accuracy.
Furthermore, in accord with the observation by Starykh et al.\cite{sss} using 
the QMC method ($\beta=8$), a low frequency secondary peak 
develops when the temperature is lowered. 

We should also observe that the way the low frequency peak  
develops is much different from the $xx$-correlation in the $XY$ case. 
There, the low frequency maximum can be traced to the fact that we chose the
non-symmetric representation (Eq. (\ref{nss})). Indeed, as the symmetric spectra
have a maximum at $\omega=0$ for all temperatures, division by 
$(1+e^{-\beta\omega})$ results in a low frequency maximum in the non-symmetric 
correlation. In contrast, in the $\Delta=1$ model,
the symmetric $S^x_{ii}(\omega)$ has a minimum at $\omega=0$ and 
a small low-frequency peak which survives upon
division by $(1+e^{-\beta\omega})$. 

The zero frequency limit 
$S^x_{ii}(\omega\rightarrow 0)=S^z_{ii}(\omega\rightarrow 0)$, relevant
for NMR experiments, seems to be roughly temperature independent.
In contrast, it increases sensibly with decreasing temperature in 
the $XY-$model.

\subsection{Heisenberg-Ising model}
\label{sgap}
\begin{figure}
\epsfxsize=12cm
\epsffile{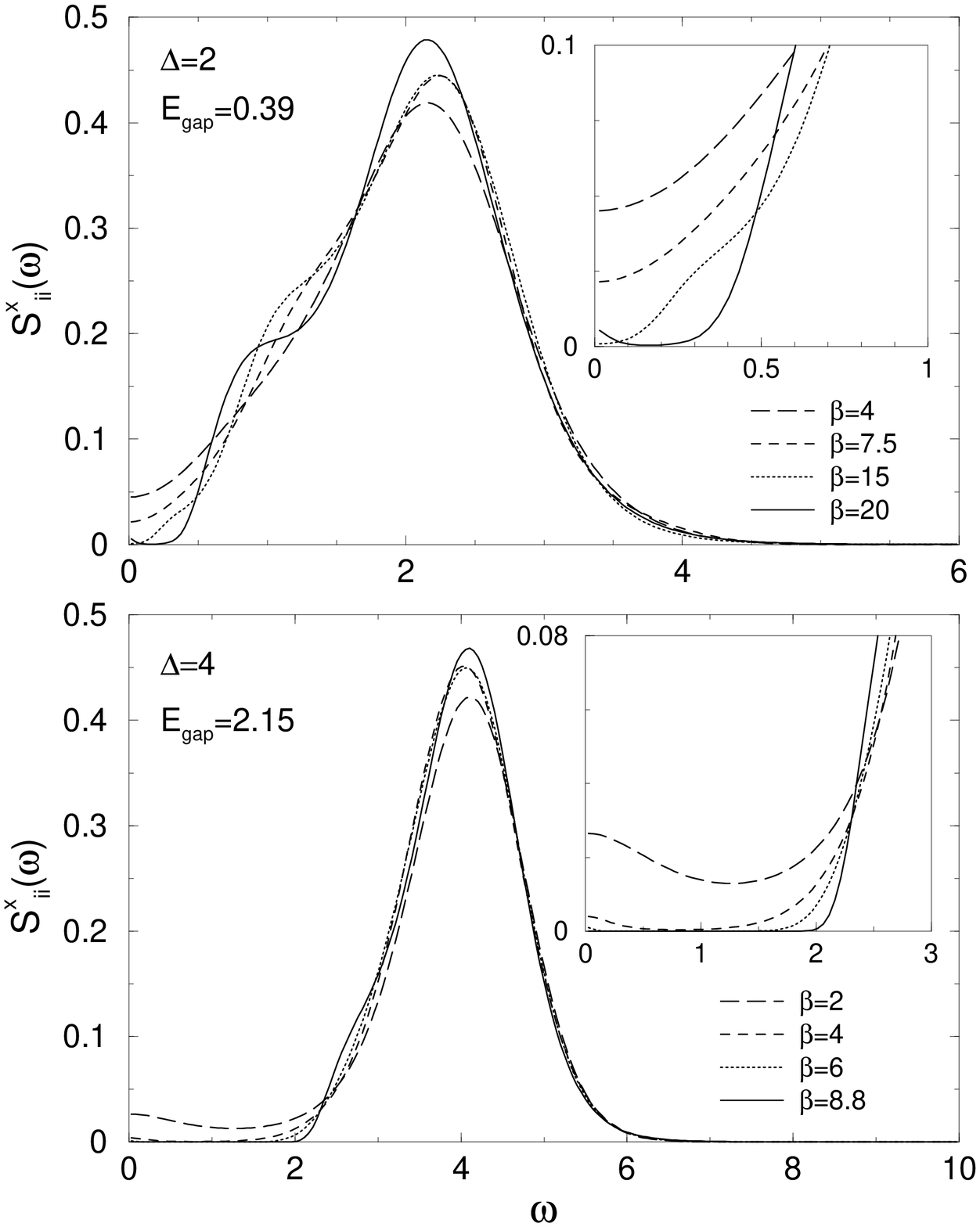}
\caption{Anisotropic model, $\Delta=2,4$.
MaxEnt from numerical data with $m=100$ and $\epsilon=0.05$.}
\label{figgap}
\end{figure}

The Heisenberg-Ising model ($\Delta>1$) is characterized by a gap
in the excitation spectrum with values $E_{gap}=0.39J$ for $\Delta=2$
and $E_{gap}=2.15J$ for $\Delta=4$.\cite{cg} 
At $T=0$, both $S^z_{ii}(\omega)$ and 
$S^x_{ii}(\omega)$ will exhibit the gap, however, a $\delta(\omega)$ function
will subsist in $S^z_{ii}(\omega)$ due to the non-vanishing matrix element
between the two degenerate ground states in the thermodynamic limit with 
total momentum quantum number $k=0,\pi$.
This matrix element was evaluated exactly by Baxter:\cite{bax}
$\langle k=0|S^z_i|k=\pi\rangle=
\frac{1}{2}\prod_{n=1}^{\infty}\left(\frac{1-q^{2n}}
{1+q^{2n}}\right)^2$ where $\Delta=\frac{q+q^{-1}}{2}$ . We have verified that
$G^z_{ii}(\beta/2)$ approaches $|\langle k=0|S^z_i|k=\pi\rangle|^2$ 
at low temperature. When reconstructing $S^z_{ii}(\omega)$, the large
weight $\delta(\omega)$ function renders the resolution at finite
frequencies poor so that the gap cannot be seen as sharply as
in $S^x_{ii}(\omega)$ shown in Fig. \ref{figgap}. Although the value of
the gap can be estimated from the spectra, the precision is not
comparable to the zero temperature DMRG method. We should also
mention that the spectrum above the gap edge seems to start 
without discontinuity, as is suggested from the two spinon 
contribution.\cite{bougou} This behavior helps in the identification 
of the gap value, in contrast to other models 
which exhibit singularities at the gap edge at $T=0$. For instance, 
the $S=1$ chain shows a square root divergence, when  
assuming a constant matrix element around the quadratic minimum of the lowest
magnon excitations at $q=\pi$. Such divergences are smeared 
out in the MaxEnt analysis.

\section{Conclusion}

We have investigated in detail whether high accuracy imaginary time 
data, obtained from the transfer matrix DMRG method, can be 
exploited in evaluating real frequency correlations after an analytical 
continuation.  
Using as a test model the the spin 1/2 Heisenberg chain and the 
MaxEnt method, we found
using the SVD and MaxEnt methods, we found that features such as the 
location of peaks and gaps can be
reliably determined. More quantitative information about 
precise lineshapes or the nature of divergences seems to lay beyond this 
procedure.

Reliable real time methods need to be developed in order to 
overcome the severe intrinsic limitations of imaginary time 
calculations.

\acknowledgments
We would like to thank A. Sandvik for useful discussions.
This work was supported by the Swiss National Foundation grant
no. 20-49486.96, the University of Fribourg and the University
of Neuch\^atel.

\begin{references}

\bibitem{tak} M. Takigawa, N. Motoyama ,H. Eisaki 
and S. Uchida Phys. Rev. Lett. {\bf 24} 4612 (1996).
\bibitem{imai} T. Imai etal., Phys. Rev. Lett. {\bf 81} 220 (1998).
\bibitem{petlan} J. Jakl\v \i c and P. Prelov\v sek Phys. Rev. B {\bf 49},
5065 (1994).
\bibitem{vismuell}V. S. Viswanath and G. M\"uller, Recursion method-Application
to Many-Body Dynamics, Lecture Notes in Physics, Vol. 23
(Springer-Verlag, New York, 1994).
\bibitem{zn} F. Naef and X. Zotos, J. Phys. C {\bf 10} L183-L190 (1998).
\bibitem{bl} M. B\"{o}hm  and H. Leschke Physica A {\bf 199} 116 (1993).
\bibitem{fab} K. Fabricius and B. McCoy, Phys. Rev. B {\bf 57}, 8340 (1998).
\bibitem{sss} O. A. Starykh, A. W. Sandvik and R. P. R. Singh,
Phys. Rev. B {\bf 55}, 14953 (1997).
\bibitem{bursill}R. J. Bursill, T. Xiang, G. A. Gehring,
J. Phys. C {\bf 8}, L583 (1996).
\bibitem{wang}X. Wang and T. Xiang, Phys. Rev. B {\bf 56}, 56 (1997).
\bibitem{Shibata} N. Shibata, J. Phys. Soc. Jpn, {\bf 66}, 2221 (1997).
\bibitem{SATSU}K. Maisinger and U. Scholl\"ock, Phys. Rev. Lett. {\bf 81}, 445 (1998).
\bibitem{shibata} N. Shibata, B. Ammon, M. Troyer, M. Sigrist, K. and Ueda
J. Phys. Soc. Jpn {\bf 67}, 1086 (1998).
\bibitem{Xiang} D. Coombes, T. Xiang, G. A. Gehring, J. Phys. C, to appear (1998); 
T. Xiang Phys. Rev. B to appear (1998).
\bibitem{Rommer} S. Eggert and S. Rommer, Phys. Rev. Lett. {\bf 81}, 1690 (1998).
\bibitem{sch2} K. Maisinger, U. Schollw\"{o}ck, S. Brehmer, H. J. Mikeska and
S. Yamamoto, Phys. Rev. B {\bf 58}, R5908 (1998).
\bibitem{sch3} S. Yamamoto, T. Fukui, K. Maisinger, 
U. Schollw\"{o}ck, J. Phys. C, (1998), to appear
\bibitem{wx} X. Wang and T. Xiang, Thermodynamics of Hubbard chains, unpublished
\bibitem{klu} A. Kl\"umper, R. Raupach and F. Sch\"onfeld, cond-mat/9809224
\bibitem{japs} A similar extension has been reported in 
T. Mutou, N. Shibata and K. Ueda, Phys. Rev. Lett. {\bf 81}, 4939 (1998).
\bibitem{karbach} M. Karbach {\it et al}., Phys. Rev. B {\bf 55}, 12510 (1997).
\bibitem{bougou} A. H. Bougourzi, M. Karbach and G. M\"uller, 
Phys. Rev. B {\bf 57}, 11429 (1998).
\bibitem{Suzuk} H.F. Trotter, Proc. Am. Math. Soc. {\bf 10}, 545 (1959); 
M.\ Suzuki, Prog. Theor. Phys. {\bf 56}, 1454 (1976).
\bibitem{Betsu} H. Betsuyaku, Prog. Theor. Phys. {\bf 73} 320 (1985).
\bibitem{white}S. R. White, Phys. Rev. Lett. {\bf 69}, 2863(1992); R. Noack and 
S. R. White, The Density Matrix renormalization group in Lecture Note in Physics,
Springer Velerg, eds. I. Peschel, X. Wang and K. Hallberg, (1998).
\bibitem{jaynes}E.T. Jaynes, in {\em Papers on Probability, Statistics and
Statistical Physics}, ed., R.D.~Rosen\-krantz, Reidel, Dordrecht,
(1983).
\bibitem{skilling}S.F.~Gull, in {\em Maximum Entropy and Bayesian Methods}, ed.
J.~Skilling, Kluwer Academic Publishers, Dordrecht, 53, 1989;
J.~Skilling,in {\em Maximum Entropy and Bayesian Methods}, ed.
P.F.~Foug\`ere, Kluwer Academic Publishers, Dordrecht, 341, 1990.
\bibitem{bretthorst}
G.L.~Bretthorst, {\em Bayesian Spectrum Analysis and Parameter
Estimation}, Springer Press, Berlin, Heidelberg, (1988) (can be
downloaded from bayes.wustl.edu).
\bibitem{wvl}W. von der Linden, Appl. Phys. A {\bf 60}, 155-165 (1995).
\bibitem{gub} M. Jarrell, J.E. Gubernatis, Phys. Rep. {\bf 269}, 133 (1996)
\bibitem{lsm} E. Lieb, T. Schulz and D. Mattis, Ann. Phys. {\bf 16},
407 (1961); B. M. McCoy, E. Barouch and D. B. Abraham, Phys. Rev. A
{\bf 4}, 2331 (1971).
\bibitem{niekat}T. Niemeijer, Physica {\bf36}, 377 (1967);
S. Katsura, T. Horigushi and M.Suzuki, Physica {\bf 46}, 67 (1970).
\bibitem{mueller}G. M\"uller and R. Shrock, Phys. Rev. B {\bf 29}, 288 (1994).
\bibitem{its}A. R. Its, A. G. Izergin, V.E. Korepin and
N.A. Slavnov, Phys. Rev. Lett. {\bf 70}, 1704 (1993).
\bibitem{bax}R. J. Baxter, J. Stat. Phys. {\bf 9}, 145 (1973).
\bibitem{cg}J. des Cloizeaux and M. Gaudin, J. Math. Phys.
{\bf 7},1384 (1966).

\end{references}
\end{document}